\documentclass[12pt, a4paper]{article}
\usepackage{cite}
\usepackage{amsmath,amssymb}
\input{colordvi.tex}
\usepackage{comment}
\usepackage{bm}
\usepackage{url}
\usepackage{ifpdf}
\ifpdf
  \usepackage{graphicx, hyperref, xcolor}     %   usepackage without driver option
\else     % For (p)LaTeX + dvipdfmx
  \usepackage[dvipdfmx]{graphicx, hyperref, xcolor}     %   usepackage with driver option
 \fi

\definecolor{rossoferrari}{HTML}{D9073D}
\definecolor{mediumblue}{HTML}{0000CD}
\hypersetup{% hyperref option list
setpagesize=false,
bookmarksnumbered=true,%
bookmarksopen=true,%
colorlinks=true,%
linkcolor=rossoferrari,
urlcolor=mediumblue,
citecolor=mediumblue,
}

\begin{document}

%%%%%%%%%%%%%%%%%%%%%%%%%%%%%%%%%%%%%%%%%%%%%%%%%%
%\begin{titlepage}

\begin{center}

\hfill ~~\\

\vskip .75in

{\Large \bf
Alternative Minimal U(1)$_{\rm B-L}$
}

\vskip .75in

{\large Kento Asai$^{(a)}$, Kazunori Nakayama$^{(a,b)}$, Shih-Yen Tseng$^{(a)}$}

\vskip 0.25in

$^{(a)}${\em Department of Physics, Faculty of Science,\\ The University of Tokyo,  Bunkyo-ku, Tokyo 113-0033, Japan}\\[.3em]
$^{(b)}${\em Kavli IPMU (WPI), The University of Tokyo,  Kashiwa, Chiba 277-8583, Japan}

\end{center}
\vskip .5in

\begin{abstract}
We provide a minimal alternative gauged U(1)$_{\rm B-L}$ model in which three right-handed neutrinos with charges of $(5,-4,-4)$ and only one ${\rm B-L}$ Higgs field with charge $1$ are introduced. Consistent active neutrino masses and mixings can be obtained if the $Z_2$ symmetry on two of the right-handed neutrinos is introduced. It predicts two heavy degenerate right-handed neutrinos, which may realize resonant leptogenesis scenario, and one relatively light sterile neutrino, which is a good dark matter candidate.
\end{abstract}

%\end{titlepage}

%\tableofcontents

\renewcommand{\thepage}{\arabic{page}}
\setcounter{page}{1}
\renewcommand{\thefootnote}{\#\arabic{footnote}}
\setcounter{footnote}{0}
%%%%%%%%%%%%%%%%%%%%%%%%%%%%%%%%%%%%%%%%%%%%%%%%%%

\newpage

%\tableofcontents

%%%%%%%%%%%%%%%%%%%%%%%%%%%%%%%%%%%%%%%%%%%%%%%%%%
\section{Introduction} \label{sec:Intro}
%%%%%%%%%%%%%%%%%%%%%%%%%%%%%%%%%%%%%%%%%%%%%%%%%%

Observed active neutrino masses and mixings are considered to be naturally explained by the seesaw mechanism~\cite{Yanagida:1979as,GellMann:1980vs,Minkowski:1977sc}.
In the seesaw mechanism right-handed neutrinos (RHNs) are introduced which are heavy and have yukawa couplings to the Standard Model (SM) leptons. After integrating out RHNs tiny active neutrino masses are obtained. RHNs not only explains the neutrino masses but also can create the cosmological baryon asymmetry through leptogenesis~\cite{Fukugita:1986hr}.

On the other hand, the existence of RHNs is required if the U(1)$_{\rm B-L}$ symmetry is gauged, in order to cancel the gauge anomaly. Conventionally three RHNs $N_i$ $(i=1,2,3)$ are introduced with the ${\rm B-L}$ charge $(-1,-1,-1)$. We call this case as ``conventional U(1)$_{\rm B-L}$''. Phenomenology of such a scenario is well studied.
However, there is another possibility for the charge assignments on three RHNs consistent with the anomaly cancellation: $(5,-4,-4)$~\cite{Montero:2007cd}. We call this case as the ``alternative U(1)$_{\rm B-L}$''. In this paper we focus on the alternative U(1)$_{\rm B-L}$.

In the alternative U(1)$_{\rm B-L}$ model, it is nontrivial to have appropriate RHN masses and yukawa couplings for the seesaw mechanism because of the unusual ${\rm B-L}$ charge assignments. In order to overcome this difficulty, several ${\rm B-L}$ Higgs fields with nontrivial charges are often introduced. Phenomenology of such alternative U(1)$_{\rm B-L}$ has been studied in, e.g. Refs.~\cite{Sanchez-Vega:2015qva,Ma:2015mjd,Nomura:2017jxb,Geng:2017foe,Singirala:2017cch,Das:2017deo,Okada:2018tgy,Das:2018tbd,Das:2019fee,Mahapatra:2020dgk}.

In this paper, we propose a alternative {\it minimal} U(1)$_{\rm B-L}$ model in which there is only one ${\rm B-L}$ Higgs field with the ${\rm B-L}$ charge $+1$. It is minimal in the sense that we do not introduce any additional field except for three RHNs and one ${\rm B-L}$ Higgs field.
The only extra assumption is that we will impose a $Z_2$ symmetry under which two RHNs are exchanged: $N_2\leftrightarrow N_3$. It drastically affects the yukawa structure and it can lead to correct active neutrino masses and mixings through the seesaw mechanism.
An interesting prediction of this model is the characteristic structure of RHN masses: two RHNs are heavy and nearly degenerate in mass and the other RHN is hierarchically light. The light RHN is stable and a candidate of dark matter (DM) of the universe.

In Sec.~\ref{sec:wo} we introduce the minimal alternative U(1)$_{\rm B-L}$ model without $Z_2$ symmetry and explain its difficulty to generate active neutrino masses. In Sec.~\ref{sec:w} we introduce a $Z_2$ symmetry and see that it solves the problem of the model without $Z_2$.
In Sec.~\ref{sec:dis} we discuss several phenomenological implications.

%%%%%%%%%%%%%%%%%%%%%%%%%%%%%%%%%
\section{Alternative minimal U(1)$_{\rm B-L}$ without $Z_2$}  \label{sec:wo}
%%%%%%%%%%%%%%%%%%%%%%%%%%%%%%%%%

First we introduce a minimal alternative U(1)$_{\rm B-L}$ without $Z_2$ and explain its basic structure. Eventually it will fail to explain the active neutrino masses, but it is instructive for later discussion.

Minimal alternative U(1)$_{\rm B-L}$ charge assignments are given in Table.~\ref{table:B-L}, where $Q_{Li}, u_{Ri}, d_{Ri}$ with $i=1$--$3$ are the SM quarks, $L_{\alpha}$, $e_{R\alpha}$ $(\alpha=e,\mu,\tau)$ are the SM leptons, $N_i$ $(i=1,2,3)$ are RHNs, $H$ is the SM Higgs doublet and $\phi$ is the ${\rm B-L}$ Higgs field. Since we introduce only one ${\rm B-L}$ Higgs field, we need higher dimensional operators to obtain yukawa couplings and some elements of RHN mass matrix. For example, $L_\alpha$ and $N_1$ can have yukawa couplings of the form $\mathcal L\sim (\phi^*/\Lambda)^6\overline L_\alpha N_1\widetilde H$ with $\Lambda$ being the cutoff scale. 
After taking the vacuum expectation value (VEV) of $\phi$, the relevant Lagrangian is given by
\begin{align}
	&\mathcal L_Y =\sum_{\alpha=e,\mu,\tau}\sum_{i=1,2,3} y_{\alpha i} \overline L_\alpha N_i \widetilde H + {\rm h.c.}, \label{LY}\\
	&\mathcal L_{N} = -\sum_{i,j=1,2,3} \frac{1}{2}M_{ij} N_i N_j + {\rm h.c.}, \label{LN}
\end{align}
where
\begin{align}
	y_{\alpha i} \sim \begin{pmatrix}
		\epsilon^6 & \epsilon^3 & \epsilon^3 \\
		\epsilon^6 & \epsilon^3 & \epsilon^3 \\
		\epsilon^6 & \epsilon^3 & \epsilon^3 
	\end{pmatrix},
	~~~~~~~~~
	M_{ij} \sim v_\phi \begin{pmatrix}
		\epsilon^{9} & 1 & 1 \\
		1 & \epsilon^7 & \epsilon^7 \\
		1 & \epsilon^7 & \epsilon^7 
	\end{pmatrix},
	\label{MN_wo}
\end{align}
omitting all $\mathcal O(1)$ coefficients. Here $\epsilon \equiv v_\phi/\Lambda$ with $v_\phi\equiv \left<|\phi|\right>$.
The detailed structure of the scalar potential $V(\phi,H)$ is not relevant for our study, but for completeness we give it here:
\begin{align}
	V(\phi,H)=\lambda_H\left(|H|^2-v_H^2\right)^2+\lambda_\phi \left(|\phi|^2-v_\phi^2\right)^2 +
	\lambda_{H\phi}\left(|H|^2-v_H^2\right)\left(|\phi|^2-v_\phi^2\right).
\end{align}
If $\lambda_H>0$ and $\lambda_\phi>0$, the potential minimum is $\left<|H|^2\right>=v_H^2$ and $\left<|\phi|\right> = v_\phi$ for $\lambda_{H\phi}^2 < 4\lambda_H \lambda_\phi$. It is possible to cure the SM Higgs vacuum instability if $\lambda_{H\phi}^2/(4\lambda_\phi) \gtrsim \lambda_H$ and $\sqrt{\lambda_\phi} v_\phi \lesssim 10^{10}\,{\rm GeV}$~\cite{Lebedev:2012zw,EliasMiro:2012ay}, although we do not go into details of this subject.

%%%%%%%%%%%%%%
\begin{table}
\begin{center}
\begin{tabular}{ |ccccc| c c c | c| c|} \hline
    $Q_{Li}$ & $u_{Ri}$ & $d_{Ri}$ & $L_\alpha$ &$e_{R\alpha}$ & $N_1$ & $N_2$ & $N_3$ & $H$ & $\phi$ \\ \hline\hline
    $1/3$ & $1/3$ & $1/3$ &$-1$ &$-1$ & $5$ & $-4$ & $-4$ & $0$ & $1$  \\ \hline
\end{tabular}
\caption{Alternative U(1)$_{\rm B-L}$ charge assignments on the SM quarks ($Q_{Li}, u_{Ri}, d_{Ri}$ with $i=1$--$3$), SM leptons ($L_\alpha, e_{R\alpha}$ with $\alpha=e,\mu,\tau$), RHNs ($N_1,N_2,N_3$), SM Higgs ($H$) and ${\rm B-L}$ Higgs ($\phi$).} 
\label{table:B-L}
\end{center}
\end{table}
%%%%%%%%%%%%%%

From the RHN mass matrix shown in (\ref{MN_wo}), one finds that the mass eigenvalues $\widetilde M_i$ $(i=1,2,3)$ and eigenstates $\widetilde N_i$ are given by
\begin{align}
	\widetilde M_1\simeq \widetilde M_2 \simeq v_\phi &~~~{\rm for}~~~\widetilde N_{1,2}\simeq \frac{1}{\sqrt 2}(N_1\pm N_{\rm H}),\\
	\widetilde M_3\simeq \epsilon^7 v_\phi &~~~{\rm for}~~~ \widetilde N_{3} \simeq N_{\rm L},
\end{align}
to the leading order in $\epsilon$, where 
\begin{align}
	N_{\rm H} \equiv \frac{1}{\sqrt 2}(N_2+N_3),~~~~~~N_{\rm L} \equiv \frac{1}{\sqrt 2}(-N_2+N_3).
\end{align}
Therefore, for $\epsilon \lesssim 0.1$, one of the RHNs becomes very light and the split mass spectrum is naturally realized. Note that $N_1$ and $N_{\rm H}$ forms a Dirac mass term and hence $\widetilde M_1$ and $\widetilde M_2$ are almost degenerate.

However, this model fails to explain the observed neutrino masses. In (\ref{MN_wo}), we see that $N_2$ and $N_3$ have yukawa couplings of similar order and they are much larger than the ones of $N_1$, only $N_{\rm L}$ dominantly contributes to the active neutrino masses after the seesaw mechanism. Thus two active neutrinos become almost massless while only one active neutrino can obtain an appropriate mass.
Actually one can show that the active neutrino masses are given by
\begin{align}
	m_{\nu} \sim \frac{v_H^2}{\epsilon v_\phi}(\epsilon^{9},\epsilon^{9},1).
\end{align}
It cannot explain the observed neutrino mass differences unless $\epsilon$ is very close to $1$, which, however, is just a limit that approaches the conventional U(1)$_{\rm B-L}$ model and not of our interest here.

%%%%%%%%%%%%%%%%%%%%%%%%%%%%%%%%%
\section{Alternative minimal U(1)$_{\rm B-L}$ with $Z_2$}  \label{sec:w}
%%%%%%%%%%%%%%%%%%%%%%%%%%%%%%%%%

In the previous section we have seen that the minimal alternative U(1)$_{\rm B-L}$ is not consistent with the observed neutrino masses. The problem is that one light RHN $(N_{\rm L})$ dominantly contributes to the seesaw mechanism and hence there appear two (almost) massless active neutrino species. As we shall see below, one can decouple $N_{\rm L}$ from the SM yukawa sector by introducing an appropriate $Z_2$ parity so that the remaining two heavy RHNs contribute to the seesaw mechanism.

Let us introduce $Z_2$ parity under which $N_2$ and $N_3$ are exchanged:
\begin{align}
	Z_2~:~ N_2 \leftrightarrow N_3.
\end{align}
In terms of the irreducible representation of $Z_2$, it is equivalent to
\begin{align}
	Z_2~:~ N_{\rm H} \rightarrow N_{\rm H},~~~~~~N_{\rm L} \rightarrow -N_{\rm L},
\end{align}
where
\begin{align}
	N_{\rm H} \equiv \frac{1}{\sqrt 2}(N_2+N_3),~~~~~~N_{\rm L} \equiv \frac{1}{\sqrt 2}(-N_2+N_3).
\end{align}

The Lagrangian is of the form (\ref{LY})--(\ref{LN}), but the structure of the yukawa and RHN mass matrix are constrained due to the $Z_2$ symmetry.
First let us see the RHN mass matrix. Due to the $Z_2$ symmetry, the mass matrix of RHNs have the structure of
\begin{align}
	M_{ij} = v_\phi \begin{pmatrix}
		c_1 \epsilon^{9} & c_2 & c_2 \\
		c_2 & c_3\epsilon^7 & c_4\epsilon^7 \\
		c_2 & c_4\epsilon^7 & c_3\epsilon^7 
	\end{pmatrix},
	\label{Mij_Z2}
\end{align}
where $c_1,c_2,c_3,c_4$ are $\mathcal O(1)$ coefficients. The mass eigenvalues of the RHNs are similar to the case without $Z_2$:
\begin{align}
	\widetilde M_1 \simeq \left(\sqrt{2c_2^2} +\frac{c_3+c_4}{2}\epsilon^7  \right)v_\phi &~~~{\rm for}~~~\widetilde N_{1}\simeq \frac{1}{\sqrt 2}(N_1+ N_{\rm H}),\\
	\widetilde M_2  \simeq \left(-\sqrt{2c_2^2} +\frac{c_3+c_4}{2}\epsilon^7  \right)v_\phi &~~~{\rm for}~~~\widetilde N_{2}\simeq \frac{1}{\sqrt 2}(N_1- N_{\rm H}),\\
	\widetilde M_3= (c_3-c_4)\epsilon^7 v_\phi &~~~{\rm for}~~~ \widetilde N_{3} = N_{\rm L}.  \label{ML}
\end{align}
For later convenience, we define $M_{\rm H1}\equiv |\widetilde M_1|$, $M_{\rm H2}\equiv |\widetilde M_2 |$ and $M_{\rm L} \equiv |\widetilde M_3|$.
Note again that $M_{\rm H1}$ and $M_{\rm H2}$ are almost degenerate: the mass difference is $\mathcal O(\epsilon^7)$. In this case, $N_{\rm L}$ is exactly the mass eigenstate while the mixing of $N_{\rm H}$ and $N_1$ are slightly modified by $\mathcal O(\epsilon^7)$ terms.
Similar to the previous case, two heavy RHNs are almost degenerate and the other one $(N_{\rm L})$ is hierarchically light.

Next let us see the yukawa sector. Due to the $Z_2$ symmetry, yukawa couplings must satisfy $y_{\alpha 2}=y_{\alpha 3} (\equiv y_{\alpha \rm H}/\sqrt{2})$. Thus the yukawa terms are written as
\begin{align}
	\mathcal L_Y =\sum_{\alpha=e,\mu,\tau}\left[ y_{\alpha 1} \overline L_\alpha N_1 \widetilde H 
	  + y_{\alpha \rm H}\overline L_\alpha N_{\rm H} \widetilde H  \right]+ {\rm h.c.}.
\end{align}
Here is the crucial observation: it is only $N_1$ and $N_{\rm H}$ that have yukawa couplings to the active neutrinos. Since both $N_1$ and $N_{\rm H}$ are heavy and have almost the same mass, it effectively reduces to the seesaw with two RHNs~\cite{Frampton:2002qc,Ibarra:2003up}. Remarkably, the light RHN, $N_{\rm L}$, decouples from the SM yukawa sector. This is due to the $Z_2$ symmetry under which only $N_{\rm L}$ flips the sign and all the remaining fields do not.
As far as the $Z_2$ symmetry is exact, $N_{\rm L}$ is stable and a good DM candidate.

\begin{figure}[t]
  \centering
  \begin{tabular}{cc}
    \includegraphics[width=0.8\hsize]{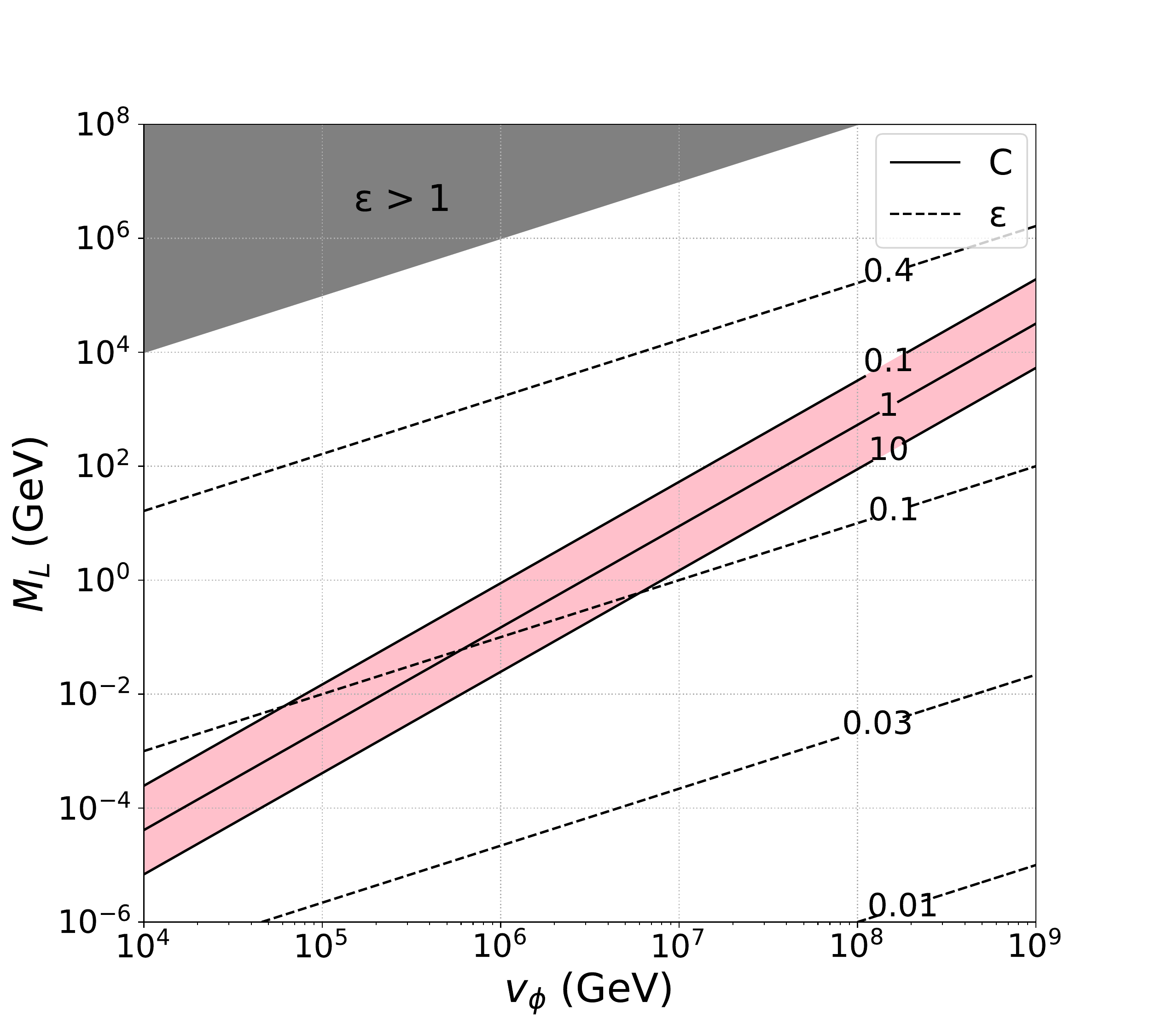}
  \end{tabular}
  \caption{The predictions on the $v_\phi$-$M_{\rm L}$ plane are shown in the plot. A band of the predicted coefficient $C$ is given in the range ($0.1, 10$) corresponding to the observed mass difference $\left | \Delta m_{23}\right | \simeq 0.05$ eV. We also present some $\epsilon$ contours and the excluded region with $\epsilon > 1$ is shaded.}
  \label{fig:C}
\end{figure}

Now the active neutrino mass matrix is obtained through the seesaw formula:
\begin{align}
	m_{\nu}=m_{\rm D}(M_{N})^{-1} m_{\rm D}^{\rm T}
	\simeq \frac{\epsilon^9 v_H^2}{\sqrt{2}c_2 v_\phi}
	\begin{pmatrix}
		2d_1d_4 & d_2d_4+d_1d_5 & d_3d_4+d_1d_6 \\
		d_2d_4+d_1d_5 & 2d_2d_5 & d_3d_5+d_2d_6 \\
		d_3d_4+d_1d_6 & d_3d_5+d_2d_6 & 2d_3d_6
	\end{pmatrix},
	\label{mnu}
\end{align}
where
\begin{align}
	m_{\rm D} = v_H \begin{pmatrix}
		d_1 \epsilon^6 & d_4\epsilon^3  \\
		d_2 \epsilon^6 & d_5\epsilon^3  \\
		d_3 \epsilon^6 & d_6\epsilon^3  
	\end{pmatrix},~~~~~~
	M_N= v_\phi \begin{pmatrix}
		c_1 \epsilon^{9} & \sqrt{2}c_2  \\
		\sqrt{2}c_2 & (c_3+c_4)\epsilon^7
	\end{pmatrix},
	\label{mDMN}
\end{align}
with $d_1$--$d_6$ being $\mathcal O(1)$ coefficients.\footnote{
	The seesaw structure is similar to those studied in Refs.~\cite{Nakayama:2016gvg,Nakayama:2017cij} in the context of sneutrino chaotic inflation.
} Note that the neutrino mass matrix is rank 2, since the matrix $M_N$ is rank 2 and there is one massless neutrino. The neutrino mass eigenvalues are given by
\begin{align}
	m_\nu \simeq  \frac{\epsilon^9 v_H^2}{\sqrt{2}c_2 v_\phi}
	\begin{pmatrix}
	0, &
	d_A+d_B, &
	d_A-d_B
	\end{pmatrix},
\end{align}
where $d_A$ and $d_B$ are $\mathcal O(1)$ coefficients defined by
\begin{align}
	d_A \equiv \sum_{i=1,2,3}d_i d_{i+3},~~~~~~
	d_B\equiv \sqrt{\left(\sum_{i=1,2,3}d_i^2\right)\left(\sum_{j=4,5,6}d_j^2\right)}.
\end{align}
Using (\ref{ML}) to eliminate $\epsilon$, the typical active neutrino mass is estimated as
\begin{align}
	m_\nu &= \left|\frac{d_A\pm d_B}{\sqrt{2}c_2(c_3-c_4)^{9/7}}\right| \frac{v_H^2 M_{\rm L}^{9/7}}{v_\phi^{16/7}}
	\equiv C_{\pm}\frac{v_H^2 M_{\rm L}^{9/7}}{v_\phi^{16/7}} \\
	&\simeq 0.01\,{\rm eV}\times C_{\pm} \left( \frac{M_{\rm L}}{10^4\,{\rm GeV}} \right)^{9/7} \left( \frac{10^9\,{\rm GeV}}{v_\phi} \right)^{16/7}.
	\label{mnu_Z2}
\end{align}
Fig.~\ref{fig:C} shows the contours of $\mathcal O(1)$ coefficient $C$ (one of the $C_{\pm}$) on the $(v_\phi,M_{\rm L})$ plane, which reproduces the observed neutrino mass difference $|\Delta m_{23}| \sim 0.05\,{\rm eV}$~\cite{Zyla:2020zbs}. The other observed neutrino mass difference $|\Delta m_{21}|$ is also reproduced if the other coefficient of $C_{\pm}$ is slightly smaller.

In calculating the active neutrino mass matrix, all the $3\times 2$ elements of the yukawa matrix equally contribute to the final expression of the neutrino mass matrix and hence there are enough degrees of freedom to reproduce the observed neutrino masses, mixings and CP phase.
Yukawa couplings (or $d_1$--$d_6$) are in general complex and hence contain $12$ real parameters. Among them, $3$ phases can be absorbed by the phase rotation of $L_\alpha$ and hence there are $9$ parameters remained. On the other hand, observable quantities in the neutrino sector are $2$ neutrino masses (since the lightest neutrino is massless), $3$ mixing angles, $1$ Dirac CP phase and $1$ Majorana phase (one more Majorana phase is not physical when one of the neutrinos is massless). Thus $9$ parameters in the model are enough to explain the observed data.

%%%%%%%%%%%%%%%%
%%%%%%%%%%%%%%%%

%%%%%%%%%%%%%%%%%%%%%%%%%%%%%%%%%%%%%%%%%%%%%%%%%%
\section{Discussion} \label{sec:dis}
%%%%%%%%%%%%%%%%%%%%%%%%%%%%%%%%%%%%%%%%%%%%%%%%%%

\paragraph{Dark matter}

In our $Z_2$ symmetric model, the light RHN ($N_{\rm L}$) is stable and a candidate of DM. Production mechanisms $N_{\rm L}$ particles are highly dependent on properties of DM and inflation models. There are various types of production scenarios, depending on the reheating temperature $T_{\rm R}$ and whether DM is in thermal bath in the early universe or not.
In the thermal freeze-in scenario~\cite{Khalil:2008kp,Kusenko:2010ik,Kaneta:2016vkq,Biswas:2016bfo}, $N_{\rm L}$ particles are produced through the scattering of SM fermions in thermal bath with the ${\rm B-L}$ gauge boson exchange. The abundance is dominated by those created at the temperature $T\sim T_{\rm R}$ ($T_{\rm R}$ denotes the reheating temperature), assuming $T_{\rm R}\gtrsim M_{\rm L}$. The abundance in terms of the density parameter is given by~\cite{Khalil:2008kp}\footnote{
	Since the ${\rm B-L}$ charge of light RHN is $-4$, the abundance is $16$ times larger than the estimate in Ref.~\cite{Khalil:2008kp}.
}
%\textcolor{blue}{The effective relativistic degrees of freedom, $g_*$, is dependent on the temperature. For $T \gtrsim 100$~GeV, $g_* \sim 100$.}
%%
\begin{align}
	\Omega_{N_{\rm L}} h^2\sim 0.5 \left( \frac{M_{\rm L}}{10^3\,{\rm GeV}} \right)\left( \frac{T_{\rm R}}{10^3\,{\rm GeV}} \right)^3 \left( \frac{10^9\,{\rm GeV}}{v_\phi} \right)^4.
	\label{ONL}
\end{align}
The ${\rm B-L}$ gauge boson mass is assumed to be much larger than the temperature.
If $T_{\rm R} \lesssim M_{\rm L}$, on the other hand, thermal production is dominated by $T\sim M_{\rm L}$ before the completion of the reheating, since there is high temperature dilute plasma and the maximum temperature is much higher than $T_{\rm R}$ and possibly than $M_{\rm L}$. The abundance in this case is given by
\begin{align}
	\Omega_{N_{\rm L}} h^2\sim 0.5 \left( \frac{10^3\,{\rm GeV}}{M_{\rm L}} \right)^3\left( \frac{T_{\rm R}}{10^3\,{\rm GeV}} \right)^7 \left( \frac{10^9\,{\rm GeV}}{v_\phi} \right)^4.
	\label{ONL_lowTR}
\end{align}
It can reproduce the observed DM abundance for reasonable choice of the reheating temperature. Here it is implicitly assumed that $N_{\rm L}$ is not thermalized, although for higher $T_{\rm R}$ and/or smaller $v_\phi$ it is possible that $N_{\rm L}$ takes part in thermal bath. In such a case the final relic abundance is determined by the $N_{\rm L}$ self-annihilation cross section through the freeze-out mechanism~\cite{Okada:2010wd}.
Another possible source is the decay of a scalar~\cite{Shaposhnikov:2006xi,Kusenko:2006rh,Petraki:2007gq,Merle:2013wta}. In our model it may be possible to obtain a correct relic abundance through the $\phi$ decay, but it strongly depends on the dynamics of $\phi$ in the early universe. The detailed discussion will be done in a future work.

\paragraph{Leptogenesis}

In our model, two heavy RHNs are almost degenerate in mass. The mass difference is of the order of $\left(M_{\rm H1}-M_{\rm H2}\right)/ M_{\rm H1} \sim \mathcal O(\epsilon^7) \sim M_{\rm L}/M_{\rm H1}$. If they are produced thermally or non-thermally, their CP-violating decay can generate a lepton asymmetry. Due to the degeneracy of two RHNs, the so-called resonant leptogenesis may occur~\cite{Covi:1996fm,Pilaftsis:1997dr,Buchmuller:1997yu,Pilaftsis:2003gt,Pilaftsis:2005rv,Anisimov:2005hr,Dev:2017wwc}. The CP asymmetry parameter per RHN decay is given by
\begin{align}
	\eta_i = \frac{{\rm Im}\left[(\widetilde y_{i\alpha} \widetilde y^\dagger_{\alpha j})^2\right]}{8\pi (\widetilde y_{i\alpha} \widetilde y^\dagger_{\alpha i})} \frac{M_{{\rm H}i}M_{{\rm H}j}(M_{{\rm H}i}^2-M_{{\rm H}j}^2)}{(M_{{\rm H}i}^2-M_{{\rm H}j}^2)^2+R^2},
\end{align}
for $i=1,2$, where $\widetilde y_{i\alpha}$ denotes the yukawa matrix in the RHN mass eigenstate basis and $R$ denotes the regulator, which is of the order of $M_{\rm H}\Gamma$ with $\Gamma$ being the RHN decay width~\cite{Garny:2011hg,Garbrecht:2011aw,Iso:2013lba,Iso:2014afa,Garbrecht:2014aga}.
As a very rough estimate, $\Gamma \sim \epsilon^6 M_{\rm H}/(8\pi)$, which is more or less the same order as $M_{\rm L}$ and hence $R\sim M_{{\rm H}i}^2-M_{{\rm H}j}^2$. Remarkably, in this model the CP asymmetry is automatically maximized. Note that typically $\widetilde y_{i\alpha} \sim \epsilon^3$, we have the CP parameter of $\eta_i\sim \mathcal O(0.1)$.

The sphaleron effect converts the lepton asymmetry into the baryon asymmetry~\cite{Kuzmin:1985mm}. The final baryon asymmetry is $Y_{\rm B} \simeq -(28/79)\kappa\sum_i \eta_i Y_{\widetilde N_i}$, where $Y_{\widetilde N_i}$ denotes the RHN number density divided by the cosmic entropy density and $\kappa$ represents the washout factor~\cite{Giudice:2003jh}. In our case, for relatively low reheating temperature $T_{\rm R}$ as seen from (\ref{ONL}), the initial RHN abundance can be highly suppressed compared with thermal abundance, which can lead to small  $Y_{\widetilde N_i}$ for reproducing the observed baryon asymmetry $Y_{\rm B}\sim 10^{-10}$. One should also be careful that the cutoff scale of the theory is $\Lambda \sim M_{\rm H}/\epsilon$, and the temperature should be lower than it for reliable calculation. We leave detailed analyses of this issue for a future work.

\paragraph{Comparison with conventional U(1)$_{\rm B-L}$}

One may introduce a $Z_2$ parity in the conventional U(1)$_{\rm B-L}$ model in which all three RHNs have ${\rm B-L}$ charge $-1$~\cite{Okada:2010wd} (see e.g. Refs.~\cite{Okada:2016gsh,Okada:2016tci,Kaneta:2016vkq,Okada:2018ktp,Okada:2020cue,Borah:2020wyc} for phenomenology of such a model). 
For example, let us assume only $N_1$ is $Z_2$ odd. Then $N_1$ is decoupled from the SM yukawa sector and becomes stable, and hence it is a DM candidate. The seesaw mechanism works with two RHNs: $N_2$ and $N_3$. In this case, there does not appear a special mass structure for three RHNs. Naturally all the three RHNs should have masses of the same order unless some additional flavor symmetry is introduced.
On the other hand, in an alternative U(1)$_{\rm B-L}$ with $Z_2$, due to the special charge assignments of ${\rm B-L}$, there appears a special RHN mass structure: two heavy degenerate RHNs and one light RHN. 
It is a characteristic feature of minimal alternative U(1)$_{\rm B-L}$ that is distinguished from conventional models.

\paragraph{Constraints}

So far we have not discussed the size of U(1)$_{\rm B-L}$ gauge coupling. If it is not very small, the ${\rm B-L}$ gauge boson mass is not far from $v_\phi$ and it is heavier than $\sim {\rm TeV}$ in the range of Fig.~\ref{fig:C}. The LHC constraint as well as other experimental constrains are safely avoided for such a heavy gauge boson~\cite{Okada:2016gsh,Okada:2016tci,Okada:2018ktp}.
If the gauge coupling is very small, on the other hand, the ${\rm B-L}$ gauge boson becomes light and there are several constraints from collider experiments and stellar physics~\cite{Kaneta:2016vkq,Okada:2020cue}, which would open up a possibility that the signal of ${\rm B-L}$ gauge boson will be detected in future.

\paragraph{Additional operator}

We started with the Lagrangian (\ref{LY}) and (\ref{LN}) and integrated out RHNs to obtain active neutrino masses. However, one can introduce the following operator by hand before the integration of RHNs,
\begin{align}
	\mathcal L_\nu = \sum_{\alpha,\beta=e,\mu,\tau}c_{\alpha\beta} \frac{\phi^{*2}}{\Lambda^3}\left(\overline L_\alpha \widetilde H \right)\left(\overline L_\beta \widetilde H \right)  + {\rm h.c.},
	\label{LW}
\end{align}
where $c_{\alpha\beta}$ are $\mathcal O(1)$ coefficients. It leads to neutrino masses of $m_\nu \sim \epsilon^3 v_H^2/v_\phi$. On the other hand, the neutrino mass that arises after the integration of RHNs is $m_\nu \sim \epsilon^{-1}v_H^2/v_\phi$ for the heaviest one, if there is no $Z_2$ symmetry ($N_2\leftrightarrow N_3$) as shown in Sec.~\ref{sec:wo}. Thus the latter one is dominant and the discussion that we need to introduce the $Z_2$ symmetry to suppress the latter contribution remains intact. Once the $Z_2$ symmetry is introduced, the neutrino mass that arises after integrating out RHNs is $m_\nu\sim \epsilon^9 v_H^2/v_\phi$ as shown in Sec.~\ref{sec:w}, which is much suppressed compared with the contribution (\ref{LW}). It is phenomenologically viable, although the relation between RHN masses and active neutrino masses is modified from the one that we discussed in the main text. In this case we have
\begin{align}
	m_\nu \sim \epsilon^3\frac{v_H^2}{v_\phi} \sim 0.2\,{\rm eV}\left( \frac{M_{\rm L}}{1\,{\rm GeV}} \right)^{3/7} \left( \frac{10^{10}\,{\rm GeV}}{v_\phi} \right)^{10/7},
\end{align}
instead of (\ref{mnu_Z2}). Although the prediction that the lightest active neutrino is massless is lost, the unique structure of the theory, i.e. the two heavy degenerate RHNs and one light stable RHN, is unchanged.

On the other hand, it is also possible that the contribution from the operator like (\ref{LW}) is subdominant. For example, let us introduce an approximate global $Z_3$ symmetry under which only the ${\rm B-L}$ Higgs field transforms like $\phi\to e^{2\pi i/3} \phi$ and all other fields are neutral. The yukawa terms (\ref{LY}) is allowed by this $Z_3$ symmetry, but the RHN mass term (\ref{LN}) as well as the operator (\ref{LW}) violate it. Thus they should be suppressed by some small parameter $\eta$ that represents the violation of $Z_3$ symmetry:
\begin{align}
	&\mathcal L_N = -\sum_{i,j=1,2,3} \frac{\eta}{2} M_{ij} N_i N_j + {\rm h.c.}, \\
	&\mathcal L_\nu =  \sum_{\alpha,\beta=e,\mu,\tau} \eta c_{\alpha\beta} \frac{\phi^{*2}}{\Lambda^3}\left(\overline L_\alpha \widetilde H \right)\left(\overline L_\beta \widetilde H \right)  + {\rm h.c.},  \label{Lnu}
\end{align}
where $M_{ij}$ is given by (\ref{Mij_Z2}). The contribution to the neutrino masses from the operator (\ref{Lnu}) is $m_\nu\sim \eta\epsilon^3 v_H^2/v_\phi$. The neutrino masses that arise after integrating out RHNs are calculated in the same way as Sec.~\ref{sec:w} just by multiplying RHN masses by a factor $\eta$: $m_\nu\sim \epsilon^9 v_H^2/(\eta v_\phi)$. Therefore, the latter contribution is dominant if $\eta\lesssim \epsilon^3$.
In such a case, the analysis of Sec.~\ref{sec:w} remains almost the same: the only change is to reinterpret $v_\phi$ appearing in Sec.~\ref{sec:w} as $\widetilde v_\phi \equiv \eta v_\phi$. For example, Eq.~(\ref{mnu_Z2}) is unchanged once $v_\phi$ is interpreted as $\widetilde v_\phi$.\footnote{
	Note that $v_\phi$ in Eqs.~(\ref{ONL}) and (\ref{ONL_lowTR}) should not be reinterpreted as $\widetilde v_\phi$.
}

%%%%%%%%%%%%%%%%%%%%%%%%%%%%%%%%%%%%%%%%%%%%
\section*{Acknowledgments}
%%%%%%%%%%%%%%%%%%%%%%%%%%%%%%%%%%%%%%%%%%%%
This work was supported by JSPS KAKENHI Grant (Nos. JP19J13812 [KA], 18K03609 [KN], 17H06359 [KN] and 20J22214 [ST]).

%%%%%%%%%%%%%%%%%%%%%%%%%%%%%%%%%%%%%%%%%%%%%%%%%%

%%%%%%%%%%%%%%%%%%%%%%%%%%%%%%%%%%%%%%%%%%%%%%%%%%

\end{document}